\documentclass[12pt]{article}
\usepackage[utf8]{inputenc}
\usepackage{graphicx}
\usepackage[a4paper, total={6in, 8in}]{geometry}
\usepackage{authblk}
\usepackage{blindtext}
\usepackage{times}
\usepackage{hyperref}
\usepackage{xurl}

\title{Data Analysis of the Responses to Professor Abigail Thompson's Statement on Mandatory Diversity Statements}
\author
{Joshua Paik,$^{1}$ Igor Rivin,$^{2}$ \\
\normalsize{$^{1}$School of Mathematics and Statistic, University of St Andrews,}\\
\normalsize{Mathematical Institute, St Andrews KY16 9SS, joshdpaik@gmail.com}\\
\normalsize{$^{2}$Department of Mathematics, Temple University}\\
\normalsize{1805 N Broad St, Philadelphia, PA 19122, United States, rivin@temple.edu}
}
\date{February 3, 2020}

\begin{document}

\maketitle

\baselineskip20pt

An opinion piece by Abigail Thompson in the Notices of the American Mathematical Society has engendered a lot of discussion, including three open letters with over 1400 signatures. We analyze the professional profiles of signatories of these three letters, and, in particular, their citation records. We find that when restricting to R1 math professors, the means of their citations and citations per year are ordered $\mu(A) < \mu(B) < \mu(C)$. The significance of these findings are validated using a one-sided permutation test.

\section*{Introduction}

A mandatory diversity statement, otherwise known as a Diversity, Equity, and Inclusion statement, is a statement required by the University of California and other schools when job applicants apply for tenure track roles ({\it 1}). These statements are used to assess candidates, alongside teaching and research statements, and can take precedence over the rest of their applications ({\it 2}). 

In November 2019, Abigail Thompson, chair of Mathematics at UC Davis and Vice President of the American Mathematical Society (AMS), published an essay in the Notices of the AMS, which criticized the usage of Mandatory Diversity Statements when hiring mathematics faculty (3). She described Diversity Statements as a ``political test" and compared it to McCarthyism.

In December 2019, a multitude of responses to Thompson's essay were published in the Notices, accumulating hundreds of signatures. One letter, Letter A ({\it 4}), titled ``The math community values a commitment to diversity," ``strongly (disagreed) with the sentiments and arguments in Dr. Thompson’s editorial", and hoped ``that the AMS will reconsider the way that it uses its power and position in the mathematics communities in these kinds of discussions." Another letter, Letter B ({\it 5}), titled ``Letter to the Editor," spoke of ``concerns about recent attempts to intimidate a voice within our mathematical community." In this letter they reference a blog post ({\it 6)}which encouraged faculty to ``advise grad-school-bound undergraduate students – especially students who are minoritized along some axis – not to apply to UC Davis." A final letter, Letter C ({\it 7}), titled ``Letter to the Notices of the AMS," criticized the usage of mandatory diversity statements, but affirmed the importance of diversity in mathematics. 

When these authors read the names of the signatories, an informal inverse negative binomial distribution took place in their heads. About every 20 names, did we recognize someone to be a signer of Letter A, every 5 names did we recognize a signer of Letter B (the second author was a signer), and every other name on Letter C. Indeed, some of the most well known mathematicians of the last 50 years - Mikhail Gromov, Nalini Anantharam, Jeff Cheeger, Cheryl Praeger, Sun-Yung Alice Chang, Anna Erschler, Noga Alon, Helmut Hofer, James Simons - and potentially less well known, but very well respected and established mathematicians, were signers of C. (An almost equally impressive list of signatories of B can be constructed.) So we came to wonder, to what extent could we quantify the observed difference in ``prestige" between signatories? 

Many ideas came to mind, but we chose to focus on bibliometrics. Of course, it should be noted and emphasized that citations and h-indices do not impose a total order on the quality of a mathematician. For example, Stephen Smale has fewer citations than Terence Tao, but it would be very difficult to distinguish who is in fact the better mathematician. Also differences between fields --- like number theory and partial differential equations --- lead to different citation counts. And really, who could accurately distinguish between solving one important problem with another important problem? However, \textit{citations generally reflect the mathematical community's opinion of a person, and are the only empirical metric of assessing merit}. More importantly, while assessing differences in citations between one person and another may be futile --- though certainly there is a difference in scientific achievement between someone with one citation and one thousand citations --- assessing the differences in citations between two groups often can be useful. It is well known that bibliometrics plays a huge part on the most important university rankings, and most rankings seem mostly accurate. 

So we proceed by analyzing the age and gender of letter signers, the number of MathSciNet citations and citations per year of R1 Math Professors who were letter signers, the number of Google Scholar citations, citation per year, and h-indices of letter signers. We prefer restricting our attentions on R1 Math Professors on MathSciNet because is a higher quality data source when comparing mathematicians. We also assess the distributions of MathSciNet and Google Scholar citations, and determine the signatories who are Fellows of the AMS.

\section*{Data and Data Collection}
Data was collected on December 16-18, 2019, and again on January 18-20 from Google Scholar, Mathematics Genealogy Project, and MathSciNet. After a list of names and affiliations were scraped from the AMS response letters on December 16, 2019, signers were searched on Google Scholar and their citation numbers and h-index were collected. This was done using the scholarly API and extensive manual checks. Then the math-genealogy-scraper ({\it 8}) was used to calculate PhD years and errors like duplicate names or dubious citations numbers were also corrected manually. MathSciNet entries were collected manually of R1 Math Professors, verifying publications and fields as necessary. Finally, the data was merged with the QSIDE dataset ({\it 9}) released on December 28th, with corrections made January 20th. After data collection was completed, we computed citations per year by first calculating an a (PhD) age, equal to 2020 $-$ the year of PhD graduation, and then citations per year by dividing citations by age. It is reasonable to suspect that the year of PhD graduation and the year of first publication are closely related.

63.96\% of the Google Scholar citations data is missing,  but this data is as complete as possible. 38.28\% of the PhD years are missing. 0\% of MathSciNet data for R1 Math Professors is missing. It should be noted that there exists discrepancies between the Notices letter accessed on December 16, 2019 and what is online now. This is because persons who wanted to add their names to the list of signatories were allowed to do so after online publication.

\section*{Data Analysis}
\subsection*{Exploratory Analysis}
Before proceeding with the analysis, we will take note of three potential issues. The first issue is the availability of data and consequently the interpretation of p-values. While 100\% of the Math Sci Net data is present for R1 Math Professors, 63.96\% of the Google Scholar data is missing for the entire data set, and 54.52\% of the Google Scholar data is missing when restricting solely to R1 Math Professors. Moreover, when looking at signatories who have Google Scholar Profiles, it certainly appears that older signatories do not have Google Scholar profiles. With this noted, we address this by setting our significance level for the interpretation of p-values at 5\% for MathSciNet citations, and 4\% for Google Scholar citations, hence widening the potential for inconclusivity. 

The second issue is multiple p-value testing. In the following sections, we will perform 27 hypothesis tests, and we will reject the null hypothesis that $\mu(population_1) = \mu(population_2)$ in favor of the alternative, $\mu(population_1) < \mu(population_2)$, 20 times. So this means that there is a $1-0.95^{20}=64.2\%$ chance that one of these rejections were a ``false positive." A Bonferroni correction would suggest that we adjust the level at which we reject our p-values to $0.05/27=0.0023$ ({\it 10}). However, we disregard this for two reasons. Firstly, multiple hypothesis testing is more an issue in fields like computational and mathematical neuroscience, where measurements are at the microscopic level and tens of thousands of hypothesis tests are performed in single experiments. When the raw metrics we use are on such large counts, when the number of hypothesis tests is relatively small (compared to other sciences), and when the adjustments to p-values above have already been made, a Bonferroni correction or other relevant correction seems unnecessary. Secondly, as we will note below, even if we make an adjustment to p-values, the most relevant hypothesis tests produce a p-value of $0.0000$, which is notably less than any adjustment to p-value. 

Finally, 6 persons in our dataset signed both letters A and B, and 72 persons signed letters B and C. These letters were not ideologically mutually exclusive, though A and C are (hence no intersection). One could be for diversity and diversity statements, but also privilege debate. One could privilege debate, and be against diversity statements. It is inconceivable that someone is both for and against diversity statements. The effect of testing non independent populations, however, is that the observed difference between two populations, in particular B and C, will be less stark.

\subsection*{Distribution of Citations}

The distribution of the Google Scholar citations of signatories appear visually exponentially distributed, post transformation by a square root. This is confirmed by an exponential qqplot and a Kolmogrov-Smirnoff test.

\begin{center}
\includegraphics[scale=0.6]{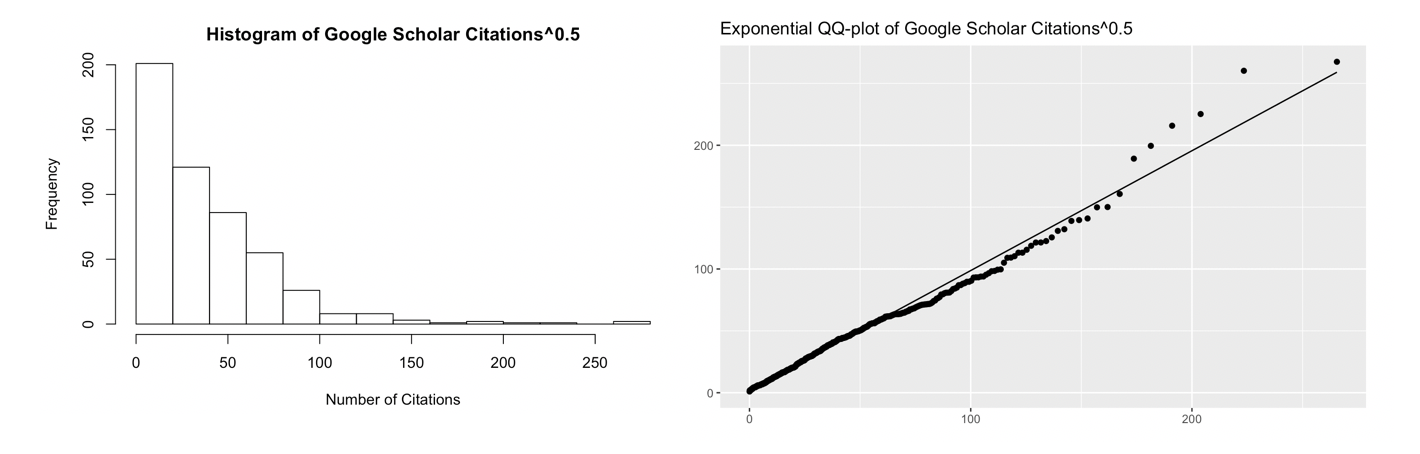}

Figure 1: Histogram and exponential qqplot of the Google Scholar citations of signatories. Considering the fit of the line, it appears the Google Scholar citations are exponentially distributed. 
\end{center}

\subsection*{Age and Citations}

Before we begin our comparison of mean citations between the signatories of letters A, B, and C, we address whether it is appropriate to do so, considering that the signatories of A are significantly younger than the signatories of B, who in turn are significantly younger than signatories of C. One may conjecture that citations or citations per year grow with age, and as a consequence any comparison is moot. 

We first plot and linearly regress the MathSciNet citations, citations/year and citations/year$^{1.3}$ over age of the R1 Math Professors who were signatories. Visually, it appears that citations grow with age, but citations/year and citations/year$^{1.3}$ do not, with regression coefficients 0.6178 (95\% confidence interval 0.115, 1.12) and 0.08284 (95\% confidence interval -0.0984, 0.2641). Increasing the year to a power of 1.3 achieves a zero in the confidence interval for the slope of age, and this fact seems invariant to other regression approaches like Theil-Sen and robust linear regression, which are less susceptible to outliers like Terence Tao. 

\begin{center}
    \includegraphics[scale=0.3]{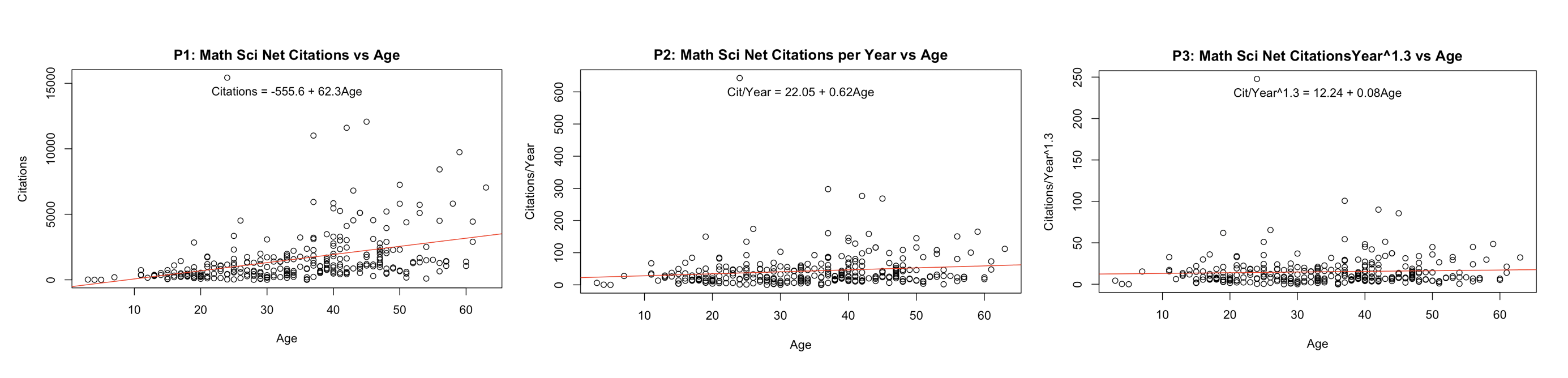}
    
Figure 2: Regression plots for citations, citations/year, and citations/year$^{1.3}$ over age. We see there to be positive correlation for citations over age, but citation/year and citations/year$^{1.3}$ over age to have less to no positive correlation. The 95\% confidence interval for slope of P2 is $(0.115,1.12)$ with adjusted $R^2 = 0.01662$ for the fit line and the 95\% confidence interval for P3 is $(-0.0984,0.2641)$ with adjusted $R^2 = -0.0006637$. So there is almost no correlation with citations/year and citations/year$^{1.3}$ with age. 
\end{center}

Of course, the art of statistics is to look at data and to make sure your eyes are not tricking you. So we produce the following cumulative mean (or median) graphs of citations over age, where each point is the mean (or median) of all signatories who are equal to or younger than the given age. At all points, signatories of letters B and C have a greater mean (or median) number of citations than signatories of letter A. 

\begin{center}
\includegraphics[scale=0.35]{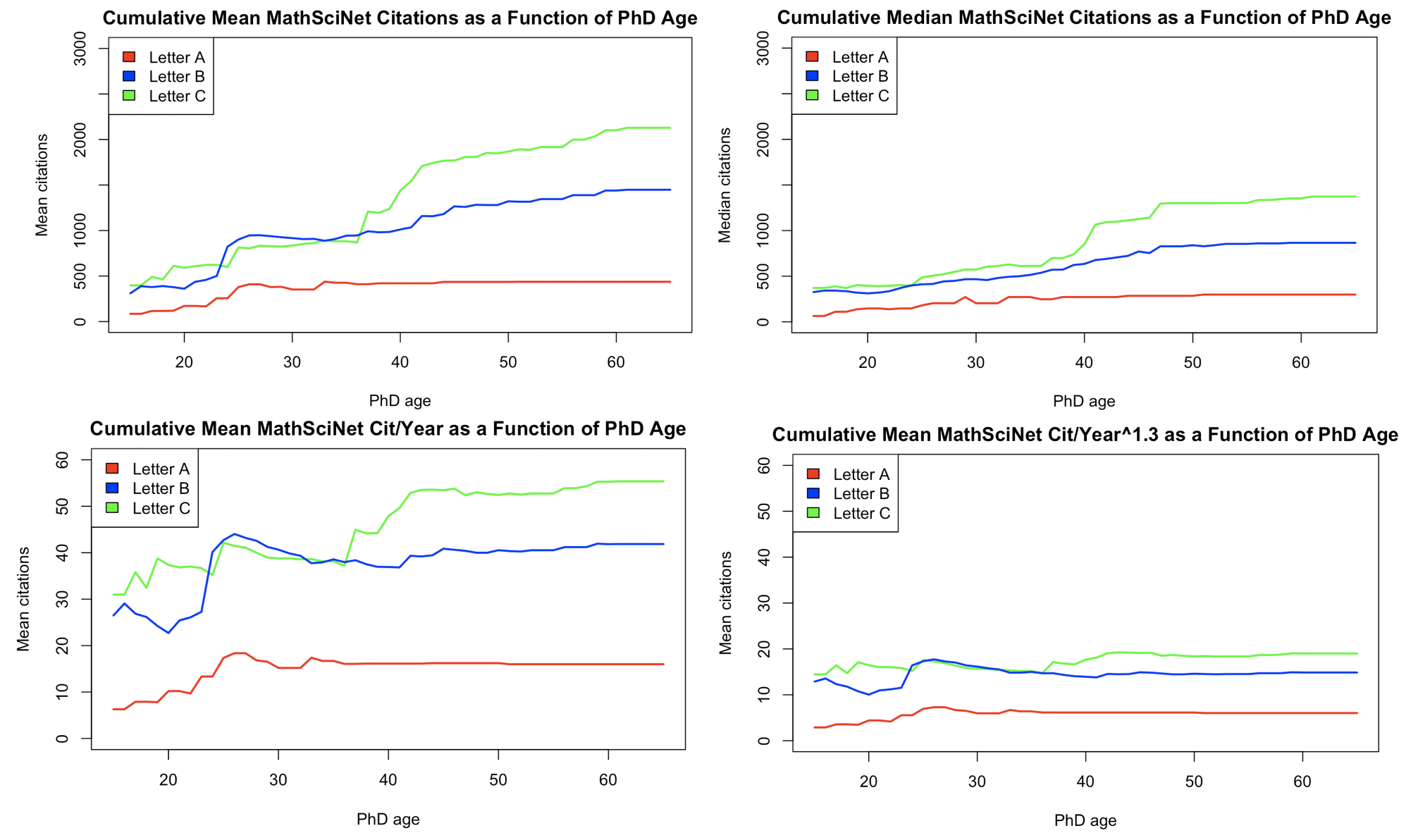}

Figure 3: Cumulative mean and median of AMS citation, cumulative mean of citations/year, and cumulative mean of citations/year$^{1.3}$ of R1 Math Professors who were signatories of letters A, B, or C. At each point in the the graph, we compute the cumulative mean and median of all signatories younger than the corresponding age.
\end{center}

Finally, we propose a final check to show that $\mu(A)<\mu(B\cup C)$. We will randomly sample a population of 20 from A, called $X$. For each member $x \in X$, we will find every person from B and C that is within a four year age interval, ($x_{age}\pm 2$) from $x$. We will randomly sample one, and induce a new population $Y$. Then we will compare the means by storing $X-Y$. We repeat this 1,000 times and induce a distribution of differences. We interpret the resulting distribution as follows: if the distribution is primarily negative, then $X < Y$. Otherwise $X > Y$. When performed, $0\%$ of the induced distribution is greater than zero, so it is totally unlikely that when comparing similarly aged professors, that their citations are greater than the citations on letter B or C.

\subsection*{Comparison of Means via Permutation Tests}

A permutation test ({\it 11}) is a non-parametric means of assessing the significance of the difference in means between two populations. Throughout this section, we will be comparing the mean citations of two populations, X and Y. We will work under the assumption that our null hypothesis is $H_0: \mu(X) = \mu(Y)$, and our alternative is $H_1: \mu(X) < \mu(Y)$. 

A permutation test works as follows. Let $X$ and $Y$ be our relevant populations, of size $n_X$ and $n_Y$. We would like to know whether we can accept that the observed difference in means was not due to chance. We record the observed difference in means as $\delta = \mu(X) - \mu(Y)$. We then take the union of our two populations, $Z = X \cup Y$, and randomly partition $Z$ into two new sets $A$ and $B$, where $|A| = n_X$ and $|B| = n_Y$. We store $\mu(X)-\mu(Y)$ and induce a distribution $D$ of potential differences and repeat the process $n=10,000$ times. We can induce the p-value, or the probability that our observed difference was due to chance by the probability $p = |\{d \in D :d\leq \delta\}|/n$. 

We report the means and medians of the relevant statistics, with their corresponding p-values in the following tables. 

\begin{center}
    \includegraphics[scale=0.7]{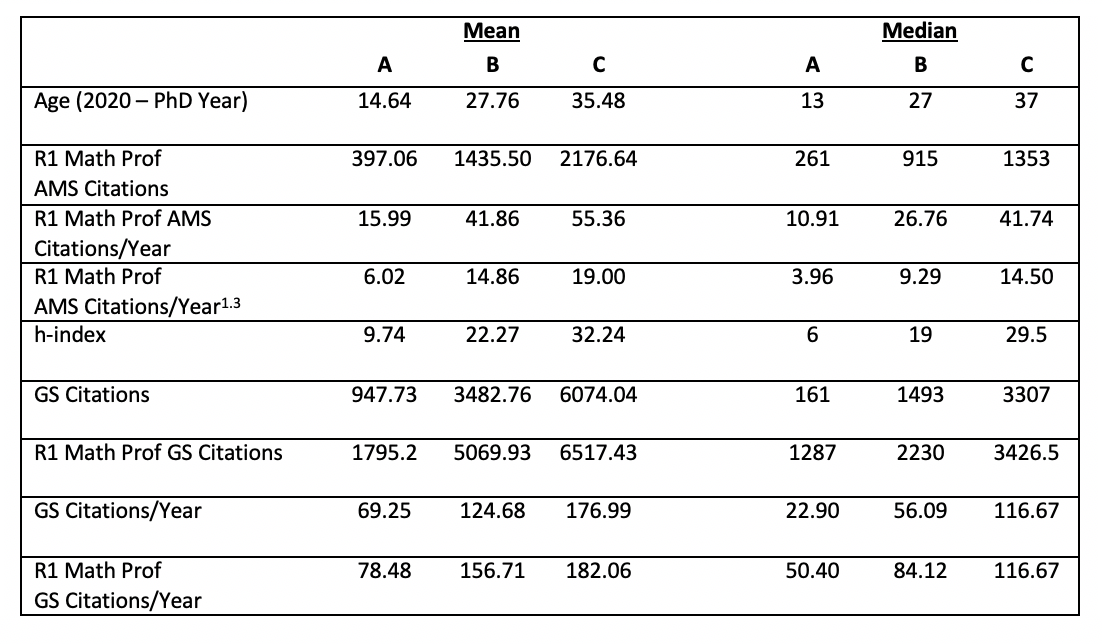}
    
    Table 1: The means and medians of the relevant statistics which label rows. In all cases, one notices the $A<B<C$ pattern. 
\end{center}
\begin{center}
    \includegraphics[scale=0.7]{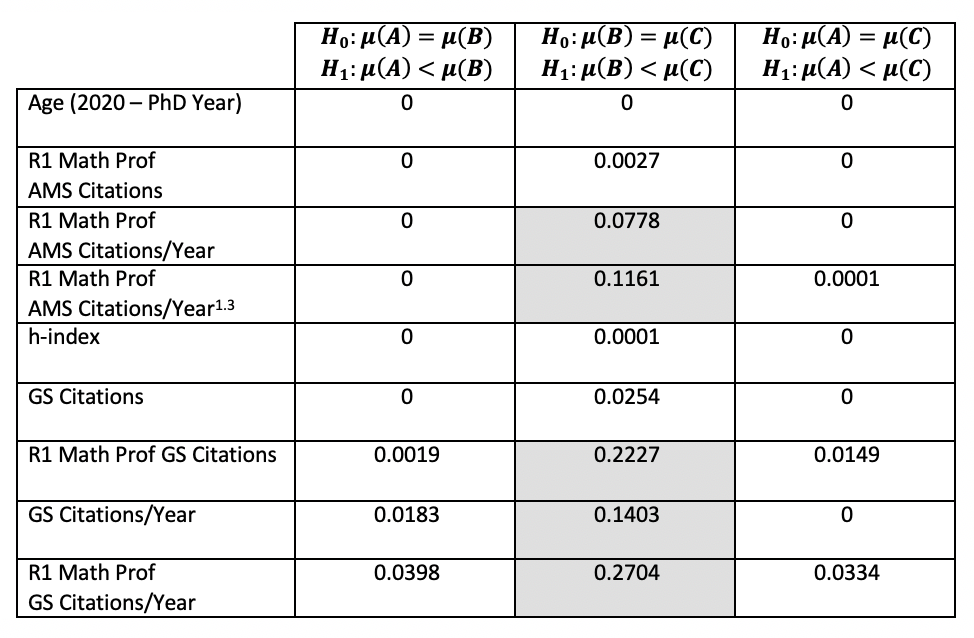}
    
    Table 2: The p-values for the hypothesis tests that label columns, for the relevant populations which label rows. All shaded values are null hypotheses which failed to be rejected in favor of the alternative at the appropriate significance level, namely 0.05 for Math Sci Net Citations and 0.04 for Google Scholar. In general, the pattern $\mu(A)<\mu(B) \leq \mu(C)$ holds. The significant overlap between signers of B and C, (n=72) leads to the null hypothesis $\mu(B) = \mu(C)$ failing to be rejected. 
\end{center}

\begin{center}
    \includegraphics[scale=0.5]{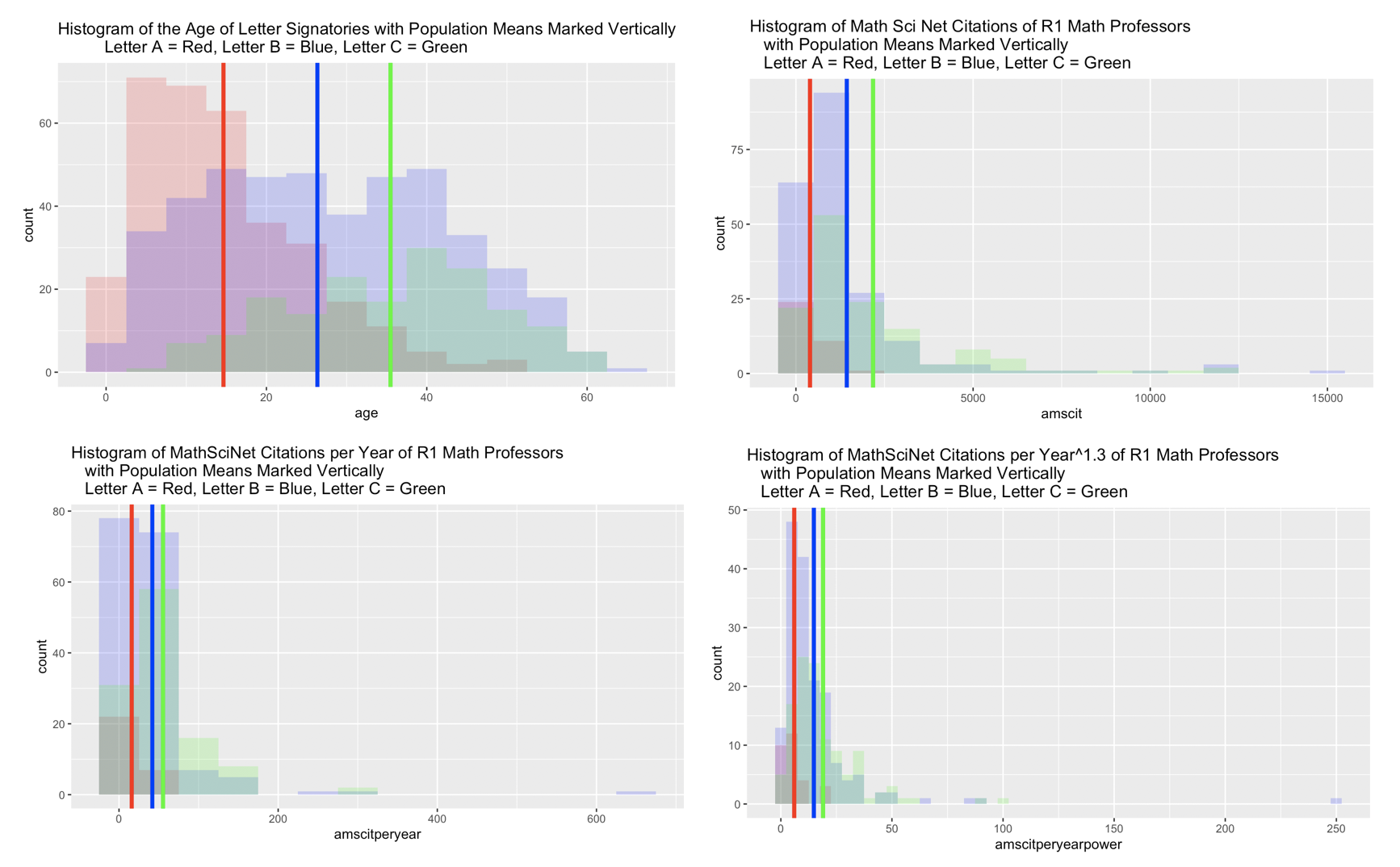}
    
Figure 4: The following plots are histograms of the distribution of the age, MathSciNet citations, citations per year, and citations per year$^{1.3}$ of signatories, with distributions of the differing populations overlaid. The vertical bars mark the population means of each distribution. 
\end{center}
\subsection*{Comparison to R1 universities and AMS fellows.}

The average AMS citations per year of R1 math professors who were signatories of letter A is 15.99, which is comparable to the University of Massachusetts - Amherst, whose math professors have an average AMS citations per year of 16.17 and a US News Ranking of 55. The average AMS citations per year of R1 math professors who were signatories of letter B is 41.86, which is comparable to the University of Minnesota, whose math professors have an average AMS citations per year of 42.07 and a US News Ranking of 19. The average AMS citations per year of R1 math professors who were signatories of Letter C is 55.36, which is comparable to the University of Chicago, whose math professors have an average AMS citations per year of 56.27, and a US News Ranking of 6.

9 signers of letter A , 136 signers of letter B, and 103 signers of letter C are fellows of the AMS. Proportionally, 1.45\% of signers of letter A, 35.8\% of signers of letter B, and 46.5\% of signers of letter C are fellows of the AMS. 

\section*{Conclusion and Discussion}

When comparing the mean number of AMS citations, citations per year, and citations per year$^{1.3}$ for R1 math professors of letter signers of A, B, and C, one finds statistically significant differences, and produces the ordering $\mu(A)<\mu(B) < \mu(C)$. These differences in mean citations are replicated with Google Scholar citations when restricted to R1 Math Professors, and produces an ordering of $\mu(A)<\mu(B) \leq \mu(C)$. To the extent that citations produce an order on merit, it seems then that the least meritorious signed A and the more meritorious signed B or C. These differences in merit, tier, and prestige persist when comparing populations to R1 Mathematics institutions relative to US News rankings, and the proportion of signatories who are fellows of the AMS. 

What explains these observed differences? One may wonder if the issue is biased sampling --- that signatories of A, B, and C would have signed another letter had they known of the other's existence. To an extent, this objection is valid. In terms of visibility, $A>B>>C$. While Letter A was by far the most visible, posted on the AMS's Inclusion/Exclusion blog with 456 Facebook and 417 Twitter shares, Letter C was restricted to select email chains and primarily targeted towards tenured professors (we were informed via personal communications). The reason for trying to restrict access to signing was to protect untenured professors from potentially damaging their careers during tenure review - one may think that such political actions could be used against them. Of course, it is not unreasonable to assume that many signatories of Letters B and C would not have signed A, just as many signatories of Letter A would not have signed B or C. 

A final concluding note. This study should not be interpreted as saying those with less citations are more likely to support diversity. The scope of this data set is not expansive or detailed enough to justify such claims. What is true is that the observed difference, in this moment in time and on this most controversial issue that has rocked the mathematical community, is that those who chose to publicly support the continued use of mandatory diversity statements had fewer citations and less merit than those who were publicly against the usage of mandatory diversity statements. 

\section*{Acknowledgements}

We are grateful to R.A. Bailey for initial discussion and helpful comments on statistical methodology. 

\section*{Data and Code}

All data and code is available at https://github.com/joshp112358/Notices.

\section*{References}

\begin{enumerate}
\item 2019-2021 Advancing Faculty Diversity:
Preliminary Report, {\it UC Office of the President; \url{https://www.ucop.edu/faculty-diversity/_files/reports/adv-fac-div-2019-21-prelim-leg-report.pdf}} (2019).
\item M. Brown, {\it Recommendations for the Use of Contributions to Diversity, Equity, and Inclusion (DEI) Statements for Academic Positions at the University of California\/} (University of California Office of the President; \url{https://academic-senate.berkeley.edu/sites/default/files/use_of_dei_statements_for_academic_positions_at_uc.pdf}) (2019).
\item A. Thompson, {\it A Word From...}, \textit{Notices of the American Mathematical Society} \textbf{66 11} (2019).
\item Various.  The math community values a commitment to diversity, \textit{Notices of the American Mathematical Society} \textbf{66 12} (2019). [online-only]
\item Various.  Letter to the Editor, \textit{Notices of the American Mathematical Society} \textbf{66 12} (2019). [online-only]
\item C. Topaz {\it Diversity statements in hiring, the American Mathematical Society, and UC Davis}: \url{https://qsideinstitute.org/2019/11/19/diversity-statements-in-hiring-the-american-mathematical-society-and-uc-davis/} (2019).
\item Various.  Letter to the Notices of the AMS, \textit{Notices of the American Mathematical Society} \textbf{66 12} (2019). [online-only]
\item J. Kun, {\it math-genealogy-scraper} \url{https://github.com/j2kun/math-genealogy-scraper}.
\item C. Topaz {\it et al.} (2020) \url{https://qsideinstitute.org/2020/01/02/math-letters-diversity-research-finalized-submitted/}
\item A. Gelman, J. Hill, M. Yajima, Why We (Usually) Don’t Have to Worry About Multiple Comparisons, {\it Journal of Research on Educational Effectiveness} \textbf{5}, 189–211 (2012).
\item P. Good, \textit{Permutation Tests}, (Springer Series in Statistics, Huntington Beach, 1994).

\end{enumerate}
\end{document}